# Voting Framework for Distributed Real-Time Ethernet based Dependable and Safe Systems

Hans Dermot Doran.
Zurich University of Applied Sciences
Institute of Embedded Systems
CH-8401-Winterthur, Switzerland
donn@zhaw.ch

*Abstract*— **In many industrial sectors such as factory automation and process control sensor redundancy is required to ensure reliable and highly-available operation. Measured values from N-redundant sensors are typically subjected to some voting scheme to determine a value which is used in further processing. In this paper we present a voting framework which allows the sensors and the voting scheme to be configured at system-configuration time. The voting scheme is designed as a Real Time Ethernet profile. We describe the structure of the voting system and the design and verification of the framework. We argue the applicability of this sub-system based on a successful prototype implementation.**

## I. Introduction

Safety and dependability in distributed industrial systems is costly to implement and certify. In many industries, system engineers have begun to move from specifying application specific components to interfacing modular pre-certified devices with appropriate communication protocols in an effort to reduce component and deployment costs. These costs remain significant and we discern pressure from industry to reduce these costs further. One path is to develop techniques to allow the use of non-certified Commercial Off the Shelf components (COTS).

In a body of work[1] we investigate, specifically for the rail and process industries, the distribution of safe and high availability operation across an array of COTS components. This body of work has been split into phases, the derivation of process variables from un-dependable distributed COTS sensors and safe and highly available operation of COTS process controllers. This paper considers the first issue.

"Distributed COTS sensors," viewed as a domain, is composed of an application specific and a generic part, the types of sensors used are specific to the application whereas the handling is generic. We therefore felt that implementing a voting framework for such inputs would help modularise and standardise the voting process and so reduce the cost of implementation and deployment. The principle requirement was that the specific sensor handling be configurable at system-integration time and the ultimate aim is suitability for use in distributed safe systems. The conception and prototype implementation are implemented with Real Time Ethernet (RTE) protocols in mind, specifically ones that support application profiles of one hue or another, otherwise we remain protocol agnostic to maximise portability.

Our contributions include expanding the conception, if not the mathematical framework, of voting to include heterogeneous sensors, large arrays of sensors, support of multiple redundancy schemes and weighted voting.

This paper describes the conception, architecture and prototype of this generic and portable voting framework for application in COTS components and is structured accordingly. This section ends with a discussion of related work. Section II describes the use case, voting model, and how we arrived at the idea of integrating the functionality in a communication profile. Section III describes the implementation details and the final section, Section IV, draws conclusions and describes further work.

### A. Related Work

The fundamental ideas behind redundancy were originally proposed by von Neumann [1], and given mathematical form by Lyons et.al [2]. Their approach to faulty devices was stringent and it is notable that, in its original conception, voting consisted solely of error masking. Much work has gone into re-assessing this stringency by, for instance, considering intermittent faults [3] and strategies for hot and cold spares [4]. [5] provides a good summary.

Voting has received much attention [f.i. 6, 7] so much so that it was considered necessary to formulate a taxonomy [8, 9]. There are few papers dealing with distributed voting systems in the domain (real time embedded) targeted by our work and for our application domain (automation, process but also relevant for automotive). [10] considers distributed systems on a macro level, that is data transferred via files. [11] considers distributed embedded systems with proprietary communication protocols. [12] considers security aspects, an area currently a work-in-progress in all (commercially available) Real Time Ethernet protocols.

There has been much work done in the area of the use of COTS in safety related applications. Most [such as 13] try to provide some sort of isolation of COTS components in case of failure. So far, despite the considerable body of work in this area, we are unaware of COTS-based architectures in use in industry today.

---

[1] We acknowledge the the sponsorship of this project by the Swiss Commission for Technology and Innovation (CTI), number 13974.1 PFES-ES with gratitude.

Apart from an early work on the issues behind safe communication on Real Time Ethernet [14] and some work with EtherCAT [15] little has been published on the issues of safe and dependable communication over real time Ethernet networks and the application of Real Time Ethernet safety protocols.

## II. CONCEPTION

### A. Voting

Component failure can be permissible and masked if the reliability of the system is not compromised and redundancy or diversity schemes are typically used to ensure this. Application of a redundancy scheme implies some form of voting, majority or otherwise. The idea of voting has now come to mean more than simple majority voting and now includes both output value calculation and output of diagnostic information. A common redundancy schemes is triple modular redundancy (TMR,) a 2 out of 3 voting (2oo3) scheme. There are applications where higher order voting schemes may be considered appropriate. Generally a 2oo3 or 3oo5 voting scheme is taken to imply that each device has equal voting rights but in higher order systems this need not be the case. In some critical systems where sensors are presumed to fail often, sensors near the measurand could be given weighting above those situated further away so any work must facilitate implementation of such a scheme. Another silent assumption is that voting is applied on homogeneous sensors but the use of diversity in sensors may hugely increase confidence in some application spaces.

### B. Model

We model the voting process as consisting of four stages as shown in Figure 1, inspired by Parhami's discussion of voting [8].

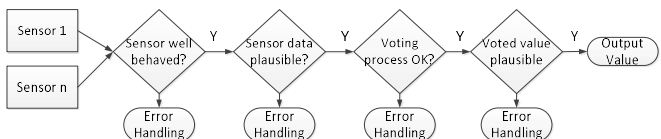

**Figure 1: Voting Process Model**

[1…n] heterogeneous sensors, all measuring the same measurand, deliver data to the first stage of the voting process where the sensors behavior is examined and a measure of acceptability determined. A well behaved remote device responds to requests promptly and transmits its data according to temporal expectations defined by the system designer. A badly behaved device can display non-desirable behaviour such as erratic transmission of frames, refusal to configuration/re-configuration demands or simply babbles uncontrollably. This behavior may be permanent or intermittent. This stage must be robust to normal system behavior such as a remote sensor being powered up before the controller, with the controller, or after the controller. Issue handling includes the detection and tracking of such faults which ultimately implies escalation to some higher (possibly human) management instance.

Once a device has been recognised as an acceptable data source the actual data can be examined and tested for plausibility. Plausibility testing may take several forms. It could be as simple as testing the data against some limits or it could take the part of grouping the data received into plausible or non-plausible categories. This test is one whose definition is in the remit of the system designer. Analogous to the first stage, if a sensors' data is considered implausible then fault handling needs to be initiated and completed.

The data voting is some algorithm that determines an output value from input data. Here also an information path to system management must be implemented to ensure that problems discovered during the voting process are reported.

Finally the value generated by the output vote may need plausibility checking.

### C. System Configuration

As we assume the number and type of sensors as well as the voting algorithms are determined at system integration time, we briefly examine a typical system configuration sequence as illustrated in Figure 2.

The application engineer will write the application and generate a run-time version together with application-orientated naming of I/O. This generated run-time version is transferred to the run-time system (controller). The I/O naming is passed to an engineering tool so that the system integrator can map the application I/O to the physically distributed I/O devices. The output generated by the engineering tool is based on (typically XML) files which describe the characteristics of the controller and the remote devices. Once the physical layout has been described the engineering tool generates a controller configuration file which is read by the controller at boot-time and used by the controller to configure the rest of the system, at run-time.

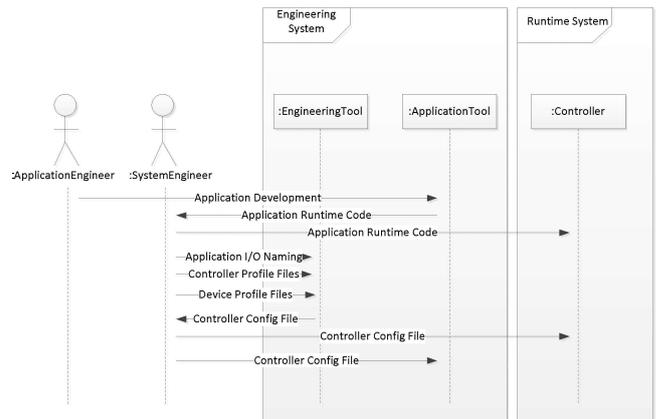

**Figure 2: Sequence Diagram Application Programming and System Integration**

A key philosophical question is whether the application programmer knows and controls the voting process or whether it is part of the system integrators domain. In other words does the application programmer code the voting or is he/she simply presented with a validated input value or appropriate error code.

In our experience there are two distinct routes to an application. One alternative is programmed in a language such as C/C++, occasionally in ADA, and run on an operating system with some real-time characteristics. An external voting framework library must then be linked into the application and possibly ported to the operating system. The second alternative is to program the application in an IEC 61131 environment. Typical IEC 61131 environments also come with engineering tools and fitted libraries of pre-defined function-blocks, from simple logic to PID controllers, which can also be supplied in the form of a safety-certified or non-certified function-block. These pre-defined blocks can be triggered by similar function-blocks originating from the application programmer and the interfaces to these function-blocks are defined and generally abstracted for the programmer.

We surmise an essential requirement is the ability to instantiate a clean and reliable interface between the voting framework and the application. As we target (RTE) distributed systems the application must interface with the communication stack. For this reason we infer that the application-communication profile interface is a suitable interface and a voting framework can be seamlessly embedded into the communication profile. The interface would be in the domain of the communication protocol with the application as a client thereof. To better illustrate this idea we shall briefly explain communication technology in the next section.

### D. Communication Techniques

In distributed automation technologies communication stacks are generally structured along the lines of the well-known OSI recommendation. The OSI recommendation stipulates layered architecture but also implies that communication conforms to the Dual-Ported RAM (DPR) model. One node writes raw-data into one side of a (local) DPR, the stack collects this data, formats it and transmits it. The stack of the remote device receives this transmission, extracts the raw data and presents it to a local DPR for reading by the application of the remote device.

The software interface to this conceptual DPR is the application layer which, abstracted to an object diagram, is shown in Figure 3. In the mid-section there are device/controller profiles and manufacturer specific profiles. Device/controller profiles describe the device or the controller - the bare minimum information a device must expose to be able to partake in structured communication. It will include essential device parameters (f.i. baud rate), software and hardware versions etc. The device will generally facilitate the idea of a manufacturer profile, a profile that is only visible to the manufacturer and used to make visible enhanced capabilities of the device available only to the manufacturer

Additionally the device will generally support an application profile, a profile that is specific to an application domain, for instance motion control, train-doors or lift control. All these profiles support some primitive object technology – the instantiation and read/write of grouping of associated parameters.

If the controller wants to modify an application-specific setting on the device it writes to the object on the controller version of the device application profile, the communication protocol transfers this value to the device's application profile. The device application may be notified of this modification by a suitable call-back or by polling and should react in an appropriate fashion to the new data.

Real time data transfers occur at specific positions in specific frames scheduled into the network bandwidth. There is generally a mapping configuration that describes the position of the application-layer objects in the real-time (Ethernet) frame.

We propose embedding the voting framework within the communication stack as a standardised profile. This proposal is supported by several important advantages. The communication stack already supports tried and tested communication and memory mapped interfaces to the application. Secondly the safe variants of the communication protocols/stacks also support the same systematic so our proposal is portable from non-safe to safe environments. Thirdly the manipulation of profiles is supported by extant engineering tools so there is no need for the development and maintenance of third party engineering workflows and associated tools.

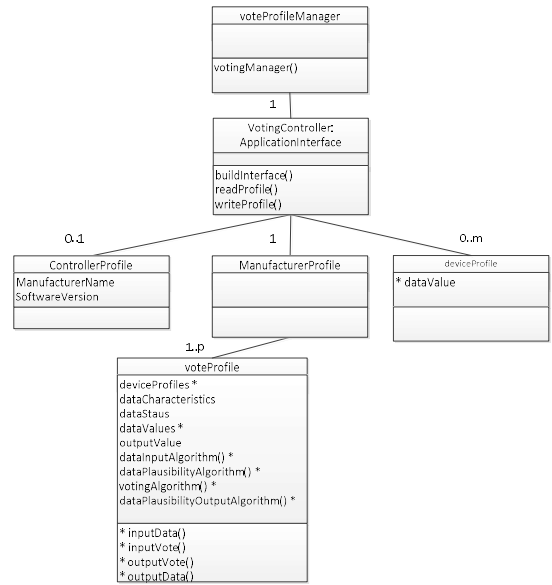

**Figure 3: Object Diagram Voting Profile Manager**

### III. IMPLEMENTATION DESIGN

#### A. Profile Design

We refer to the fundamental design in Figure 3. At system-boot, the manufacturer profile instantiates [1..p] `voteProfiles`, each `voteProfile` representing a set of devices whose delivered `dataValue` will be mapped into a list of `dataValues` voted on and the result of voting written into a single `outputValue` in the scope of the voting profile.

The profile has a number of `dataCharacteristics`, implemented as objects, for instance the maximum number of contributing devices, the type (bit-size) of data the devices contribute, etc. The profile also contains objects referencing the set of methods used for voting on the data. All profile objects can be manipulated by standard read/write/modify methods supplied by the communication stack.

Some sensors may fail completely during operation and not be replaced so inputs to the profile methods are lists of data structures and allow for management of operationally unknown (but maximum bounded) numbers of redundant sensors. At a system-specific temporal offset from the start of the application cycle the `votingManager` method is called which manages the entire voting process by calling the voting methods: `inputData`, `inputVote`, `outputVote` and `outputData` each corresponding to one of the four stages in Figure 1. In essence a method from one voting stage receives one list of data structures from the previous and divides it into two lists of "good" and "bad" sensors/data, one of which may be empty. The "bad" sensor/data lists are forwarded to error management. If a sensor behaves badly after a period on the "bad" list it may be consigned to an "unusable" list and maintenance personnel must take further action, it may equally, over time and good behaviour, "rehabilitate" itself and contribute to the system.

*B. Prototype*

For a test application three laser distance measuring sensors from di-Sorio (LHT 9-45 M 10 P3IU-B4) were each connected to a DNP9200 DILNetPC module from SSV mounted on a board developed at our institute. The DILNetPC is an Atmel AT91RM9200 32-bit ARM9 MCU with 180 MHz Clock Speed running Linux 2.26.8. This sensor is polled (UDP) by the application at the application cycle time, 20 ms. The application runs on a WAGO-I/O-IPC-C6 industrial PC (IPC) @ 600MHz using Linux as an operating system (2.6.29 incl. RT-Preempt) and the Completely Fair Scheduler CFS. The implementation language is C and the objects were implemented where appropriate as functions or as structures. The voting framework was implemented 1:1 to the design as explained above. It was tested exhaustively using CUnit testing, achieved by passing pre-defined data structures into the framework and testing against the expected control flow.

The framework was then instantiated with appropriate algorithms. For the prototype it was decided to use a common median voting technique bounded by a 2oo3 methodology. A demonstrator is extant.

I. CONCLUSIONS AND FURTHER WORK

In this paper we have described the conception and architecture of a voting framework as well as a prototype implementation for distributed systems in terms of a general proof-of-concept. We see the results achieved as a prototype pattern for the implementation of voting in distributed systems.

Our prototype application featured a cycle time of 20 ms and the real-time response of the voting system has not yet been fully determined. Our prototype also used a version of the soft-PLC, CoDeSys, running on the controllers, this version did not feature a RTE protocol interface and voted data was read by a C-function called from inside the soft-PLC environment. The limitations of these platforms interfaces to RTE protocols have yet to be plumbed.

Whilst the general mechanics are considered useful for implementation on a safe communication protocol these are difficult from an engineering point of view so we anticipate substantial work before the voting framework can be deployed in a functionally safe industry environment.